# Exploiting femtosecond laser exposure for additive and subtractive fabrication of functional materials: A Route to designer 3D Magnetic Nanostructures

*Arjen van den Berg[1], Mylène Caruel[2], Matthew Hunt[1] and Sam Ladak[1*]*

1. School of Physics and Astronomy, Cardiff University, Cardiff, CF24 3AA, UK

2. Institut National des Sciences Appliquées (INSA) of Toulouse, 35 Avenue de Rangueil, 31400 Toulouse, France ;

* LadakS@cardiff.ac.uk

**ABSTRACT**

Three-dimensional nanostructured functional materials are important systems, allowing new means to intricately control electromagnetic properties. A key problem is realising a 3D printing methodology upon the nanoscale that can yield a range of functional materials. In this letter, it is shown that two-photon lithography when combined with femtosecond machining of sacrificial layers, can be used to realise such a vision and produce 3D functional nanomaterials of complex geometry. This is demonstrated by fabricating 3D magnetic nanowires that exhibit controlled domain wall injection and propagation. Secondly, we fabricate large scale 3D artificial spin-ice structures whose complex switching can be probed using optical magnetometry. We show that by careful analysis of the magneto-optical Kerr effect signal and by comparison with micro-magnetic simulations, depth dependent switching information can be obtained from the 3DASI lattice. The work paves the way to new materials, which exploit additional physics provided by non-trivial 3D geometries.

Three-dimensional (3D) printing of materials is proving to be an important technology, allowing rapid prototyping and manufacture of a range of physical systems and across different length scales. Recent work has demonstrated multi-nozzle 3D printing in order to realise a number of multi-material systems including robot walkers and intricate origami structures upon the macroscale[1]. Reduction of such processes to physically important length scales, particularly upon the nanoscale[2], enables the investigation of a range of exotic phenomena and associated applications. Examples include 3D nanostructured systems with outstanding mechanical properties[3,4], systems with bespoke phononic bandgaps[5], frustrated magnetic nanowire lattices[6], controlled Hall coefficients in 3D chainmail metamaterials[7] and 3D conductive nanowires[8] allowing the production of coils and on-chip interconnects. There is a diverse range of methodologies available for 3D printing upon the nanoscale, but each is usually tailored to producing a limited type of material, making the realisation of more complex systems which rely upon multiple functional materials difficult[2].

A methodology to cast the shell of a polymer-based 3D nanostructure into other functional materials would be an important development in materials science, allowing the production of bespoke multifunctional materials. In order to obtain maximum benefit from such an approach, it should be compatible with conventional deposition processes allowing access to a wide range of metallic and dielectric materials with different functionalities. A key challenge with such an approach is ensuring the desired material only covers the relevant 3D nanostructure, leaving the rest of the substrate clean. A previous approach to solve this problem has used simple resist stencils[9], but this has limited applicability for 3D geometries. An alternative approach is to make use of sacrificial layers and combine this with two-photon lithography (TPL)[10].

The emerging field of 3D nanomagnetism[6,10-14] is an excellent example of how nanoscale control of 3D geometry can yield a range of new physical phenomena and advanced device concepts. Of particular note is the pioneering concept of magnetic racetrack systems[15,16], which utilise domain walls within 3D nanowire systems in order to store information. 3D magnetic nanowires were recently realised in simple angled geometries using focussed electron beam lithography of Pt, followed by $Ni_{81}Fe_{19}$ evaporation[17]. A key problem with this approach was the presence of $Ni_{81}Fe_{19}$ sheet film upon the substrate, making it more difficult to probe the magnetism of the nanowires. This obstacle was surpassed by an innovative dark-field optical magnetometry technique, which allowed the switching of such wires to be determined. An alternative approach is to remove the underlying sheet film, allowing the probe of more complex geometries.

In this communication, we outline a new methodology for realising 3D nanostructures within functional materials. We combine two powerful approaches; that of femtosecond precision machining with 3D lithography in order to realise 3D nanostructures cast into functional materials. Its power is demonstrated by fabricating lone magnetic nanowires raised above the substrate and with domain wall nucleation pads, opening up new avenues for fundamental study of 3D magnetic nanowires. Furthermore, we demonstrate that the technique can be harnessed to study magnetometry in more complex 3DASI structures taking a diamond lattice geometry.

A key challenge in realising such a process is identifying a sacrificial layer that can be ablated at high-resolution with femtosecond laser exposure, that remains chemically stable during the TPL development process and can then be finally removed in such a way that the 3D structures remain intact. We identified polyacrylic acid (PAA) as satisfying all of these criteria. Crucially, this high molecular weight polymer (Figure 1a inset) is chemically stable in the presence of commonly used solvents in lithography such as PGEMA and isopropanol, but swells when immersed in water[18]. Figure 1(a-h) shows an overview of the fabrication process. The sacrificial layer, PAA, is first spun onto a glass substrate (Figure 1b), after which trenches of the material are ablated away using high-resolution femtosecond laser machining (Figure 1c). Here it is of utmost importance that the ablation removes all polymer from the substrate, for later anchoring of a 3D nanostructure. After ablation, a negative tone photoresist is drop cast upon the sample (Figure 1d), and the desired structure is written into the ablated trenches using TPL (Figure 1e). Here the versatility of TPL can be used to produce any 3D geometry within the trench. The unexposed resist is removed by placing the sample in a propylene glycol monomethyl ether acetate (PGMEA) bath for 20 minutes, followed by an isopropyl alcohol (IPA) bath (Figure 1f). Magnetic material ($Ni_{81}Fe_{19}$) is deposited using thermal evaporation (Figure 1g) after which the PAA layer is lifted off in an aqueous sodium hydroxide (0.5 mol/l) bath (Figure 1h). A critical requirement for such a process is well-controlled ablation of the PAA sacrificial layer. Our femtosecond laser ($\tau$=100 fs, rep rate 80 MHz) operates in the infrared ($\lambda$=780nm) meaning there is insufficient energy to yield direct photolysis of carbon-based bonds within PAA. However, recent work has suggested that ultrafast lasers within the infrared can ablate polymers through multiphoton absorption[19].

To characterise the ablation process, single-voxel trenches were ablated into a PAA layer with varying laser power and scan speed. The profile of each trench was measured using atomic force microscopy (AFM). A representative AFM image is shown in Figure 2a with a 3D rendered view of the data shown in Figure 2b. A height profile of a typical trench is shown in Figure 2c. Shoulders on either side of the ablated region indicate that the ablation process is photothermal with laser exposure creating a melt pool as the PAA is heated. Subsequent exposure yields explosive vapourisation of the molten material. The recoil results in the molten material being ejected and redeposited on surrounding areas[20]. The onset of uniform ablated lines starts at a power of approximately 50mW, yielding feature sizes of below 300nm, after which it increases with laser power (Figure 2d). Below 50mW, ablated lines appear non-uniform and could not be used to produce clear trenches for this process (Figure S1). Depth of the ablated trench is shown in Figure 2e. Here, it can be seen that the depth increases for low powers before saturating at the PAA film thickness.

To demonstrate the initial feasibility of our approach, we fabricate simple $Ni_{81}Fe_{19}$ nanowires, and nanowires with domain wall nucleation pads. In both cases, the nanowire is raised 3μm above the substrate. A scanning electron microscopy image of a nanowire with nucleation pad is shown in Figure 3a. The fabricated wires have a length of 100 μm and a width of 300nm. The nucleation pad measured 20 μm × 20 μm. Elemental maps were captured using energy dispersive X-ray (EDX) analysis for Ni and Fe. Figure 3b shows spectral data from a linescan across the device showing clear L series peaks for Ni and Fe when focused on the wire. Ni and Fe line profiles are indicated with red and blue lines respectively, which show these elements upon the structure while none is detected on the substrate within the noise floor of the instrument. The magenta line indicates the energy spectrum captured from the wire with further detail shown in the upper inset. We note a peak at ~1.05 keV, away from the wire, which corresponds to the K series lines of Sodium. This signal is attributed to small deposition from the lift-off procedure using NaOH.

Magneto-optical Kerr effect (MOKE) magnetometry was used to characterise the switching in each of the nanowires. An essential requirement in the production of such systems is the realisation of isolated 3D ferromagnetic nanowires, which switch via domain wall motion and with negligible background thin-film upon the substrate. Figure 3c shows a magneto-optical Kerr effect magnetometry loop obtained upon a single magnetic nanowire. We observe a square loop with coercivity 9mT, indicative of domain wall motion. The use of soft injection pads has previously been used in 2D nanowires in order to inject domain walls controllably, but this has not been explored for 3D systems. Figure 3d and Figure 3e shows MOKE loops obtained on a wire with a 3D nanostructured injection pad. Measurement of the pad yields a rounded loop with coercivity of ~0.5mT (Figure 3d). Moving the laser spot to the opposite side of the wire yields a loop with sharp transition and coercivity of just 1mT (Figure 3e), demonstrating controlled DW injection into the nanowire. Moving the laser spot away from the structures yields no magnetic signal (Figure S2) further indicating the lift-off procedure has been successful.

With initial proof-of-principle demonstrated we move on to extend our methodology to more complex 3D artificial spin-ice (3DASI) structures which take the form of a diamond lattice geometry[21,22,23]. A key challenge in 3DASI systems is the ability to separate switching occurring upon different sub-lattices, positioned at different depths on the lattice. Due to the geometry of a diamond bond lattice, with projections of nanowires laying along perpendicular axes, one potential approach to study depth dependent switching, close to the surface, is by harnessing MOKE. Figure 4(a) shows a schematic of a 3DASI diamond-bond lattice unit cell, with nanowire sub-lattices highlighted by colour. In our experimental MOKE geometry, the wavevector is incident at an angle of ~45° with respect to the substrate, with projection along the L1 sub-lattice. There are three distinct MOKE effects which can measure different magnetisation components with respect to the incident k-vector[24] and in our experimental geometry, each of these are present as depicted in Fig 4a inset. When using an s-polarised incident beam, a longitudinal and polar signal will be present for low analyser angles, which should vanish at an analyser angle of 90°. When harnessing p-polarised light, a low analyser angle again yields a signal where the longitudinal and polar responses are dominant. However, increasing the analyser angle to 90° now isolates the transverse MOKE signal which yields only a change in intensity and in our experimental geometry should be primarily sensitive to the magnetisation component along the L2 sub-lattice. Hence in principle, one can obtain a detailed picture of the lattice switching by looking at these limiting cases in s- and p- polarised light.

A large 180μm x 180μm area of PAA was ablated away and a diamond-bond lattice of nanowires was written, subject to $Ni_{81}Fe_{19}$ deposition and lift-off. Figure 4b shows a scanning electron micrograph of the resultant structure, with the various sub-lattices labelled. Prior work showed optical magnetometry with signals originating from the sheet film occurring beyond the boundaries of the lattice[21]. Specifically, a strong dip was noted at low field at a coercivity close to the $Ni_{81}Fe_{19}$ sheet film. Such signals are very detrimental to the study of 3DASI switching, since they mask more subtle variations in signal present at low field.

With background film removed, we now carry out a preliminary exploration of the switching within this 3DASI system. Figure 4c shows a hysteresis loop obtained in s-polarisation with the field applied in the plane of the substrate, along the projection of the L1 lattice, and with the analyser angle set to ~3 degrees from extinction. No evidence of sheet film switching at low field is observed, demonstrating our sacrificial layer process has again been successful. With this milestone in place, we proceed to p-polarised light to determine if longitudinal and transverse signals now provide a means to probe sub-lattices of different depths. At low analyser angle, the measured loop closely resembles what is measured in s-polarisation as expected (Figure S3). Moving to an analyser angle of 90° provides a loop of very different shape and a sharp transition at approximately 35mT (Fig 4d).

In order to understand the features within the optical magnetometry, micromagnetic simulations were carried out whereby bipod structures were simulated in the experimental field geometries. We note that since the simulations were carried out at zero Kelvin, a systematic difference in coercivity is observed when compared to experimental results, as observed previously[25]. We consider both upper bipods found at the tops of L1 (Fig 5a, 5b) and lower bipods found at the L1/L2 intersection (Fig 5c, 5d). In each case, every vector component of the magnetisation is plotted. For the experimental data taken at an analyser angle of 3° for p-polarised light (Fig 4c), it is clear that the signal is dominated by the longitudinal component of the uppermost sub-lattice (L1), both of which are reproduced by the simulations in Fig 5a ($L_A$) and 5c (Lc). We note that the strong rounding in the experimental loop, pre-transition (Fig 4c, Red arrows) is not captured in $L_A$ or $L_C$. This suggests that a small contribution from the signal is originating from the L2 sub-lattice, with a hard-axis like characteristic as depicted in Fig 5b ($L_B$). Next, we move onto to interpret the transverse experimental loop taken in p-polarised light and at an analyser angle of 90 degrees (Fig 4d). We note immediately, that the subtle antisymmetric features (Fig 4b, Green arrows) are reproduced by the transverse component with field along long-axis as depicted in Fig 5a ($T_A$) and 5c ($T_C$). This suggests that this signal is coming from L1 and analysis of the simulated magnetisation profiles suggests it originates from magnetisation rotation, at the vertex areas (Figure S4). This rotation is removed after nucleation of a domain wall, supported by the fact that these asymmetric features occur close to the longitudinal switching field seen in Fig 4c.

The higher field transition seen in Fig 4d (purple arrows), is not explained easily by any component in the simulations depicted in Fig 5. The polar loops, ($P_B$ in Fig 5b and $P_D$ in Fig5d), show some transitions at higher fields but we note that the polar MOKE effect is not present at an analyser angle of 90°. In addition, we note that the absolute change in $M_z$ in such transitions is of order $10^{-4}$ (Figure S5). Together, these points make it unlikely that the polar MOKE effect is yielding these transitions.

We note that within our experimental setup, a small angular (<10°) misalignment of the field with respect to the L1 projection is possible. Standard models for simple single domain, high-aspect ratio systems[26,27] predict an increased switching field for angles close to 90°. Whilst we note that our 3D system is more complex in geometry, such a dependence would suggest that the transition measured in Fig 4d originates from the L2 sub-lattice. To investigate this, micro-magnetic simulations were carried out upon bipod structures, but with a 10° misalignment of the field (Figure S6). These show an approximate factor of 2 difference in switching fields for two perpendicular oriented sub-lattices, consistent with our experimental data. Overall, our preliminary analysis shows that MOKE can be exploited to obtain depth-dependent switching upon 3DASI lattices. Such work could be further refined in order to study thermal ordering in 3DASI systems.

In conclusion, we have demonstrated a novel methodology that utilises 3D lithography with a sacrificial layer in order to cast a complex 3D geometry into a functional material. We demonstrate the power of this approach by fabricating 3D magnetic nanowires with injection pads showing controlled nucleation of domain walls that can be probed optically. Going beyond this, we show that our technique can be used with complex 3DASI systems and have used optical magnetometry to observe switching upon two sub-lattices of different depth. We envisage our fabrication methodology will have an impact upon materials science and physics, allowing the influence of complex 3D geometry upon electronic, optical and magnetic properties to be investigated, paving the way to multifunctional 3D nanostructured chips.

**Acknowledgements**

SL acknowledges funding from the Engineering and Physics Research Council (EP/R009147/1) and from the Leverhulme Trust (RPG-2021-139).

**Supporting Information**

The following file is available free of charge: Supporting information

Information regarding experimental details

Figure S1: Examples of incomplete ablation.

Figure S2: Kerr signal when laser spot was moved away from nanowire.

Figure S3: Longitudinal Kerr signal for 3DASI for p-polarised light.

Figure S4: Micro-magnetic simulations of bipod showing transverse component changing at vertex.

Figure S5: Micro-magnetic simulations of bipod showing polar component.

Figure S6: Micro-magnetic simulations of bipod with 10 degree field misalignment.

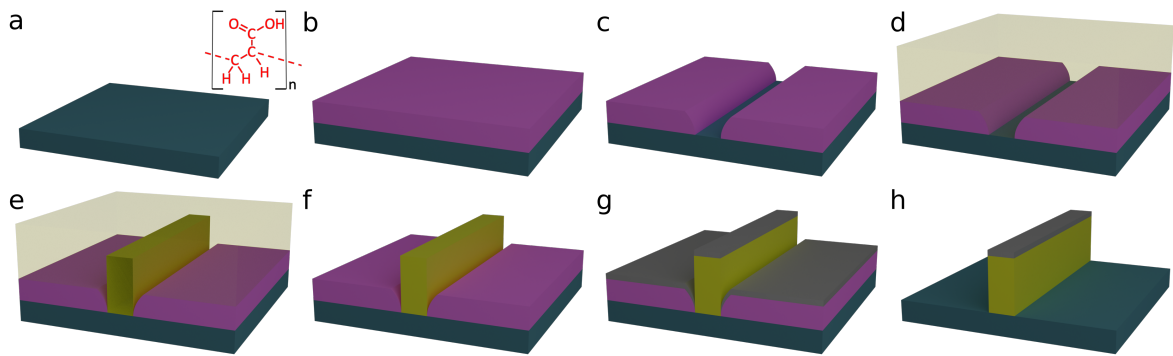

**Figure 1.** Outline of the Fabrication process. (a) A glass/ITO substrate is cleaned with acetone followed by IPA (b) Polyacryllic acid is spin coat onto the substrate. (c) Femtosecond laser machining is used to ablate the desired geometry into the PAA. (d) A negative tone photoresist is drop cast upon the sample. (e) The desired 3D geometry is written into the polymer using two-photon lithography. (f) Unexposed resist is washed away using PGMEA. (g) $Ni_{81}Fe_{19}$ is deposited by evaporation. (h) The PAA is removed using deionized water and the substrate film is lifted off.

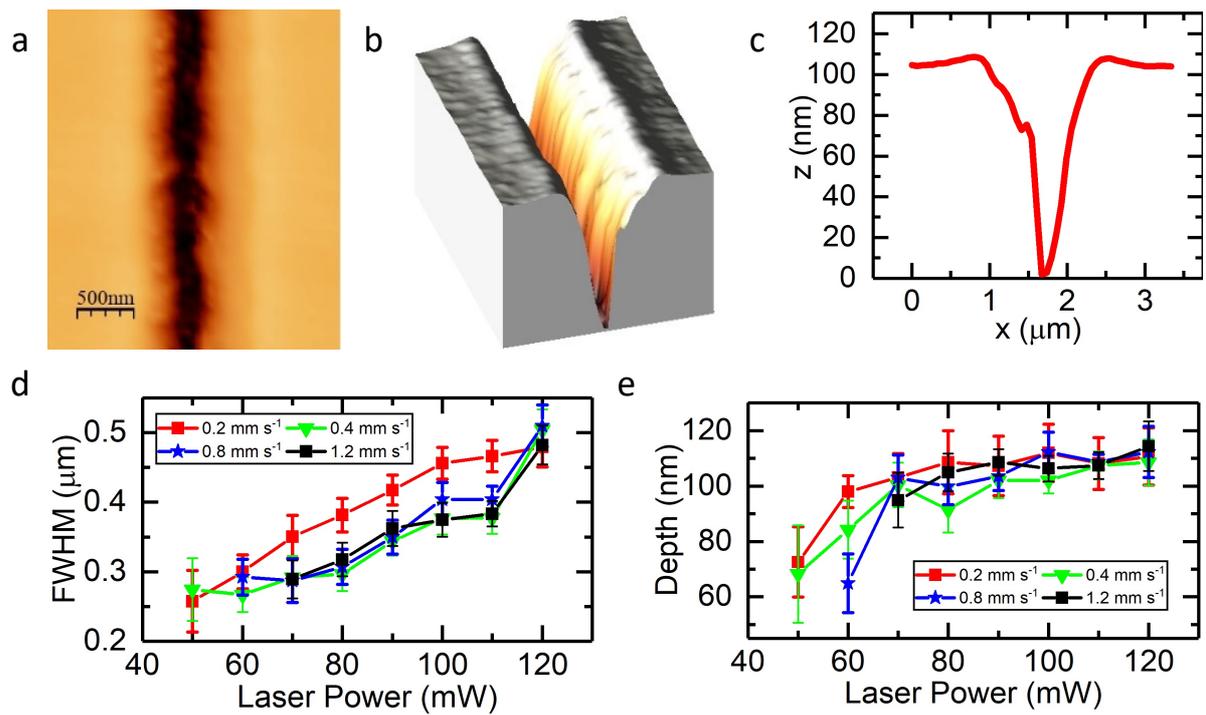

**Figure 2.** (a) Atomic force microscopy image of a single trench ablated into a polyaryllic acid layer. (b) 3D representation of ablated trench. (c) Height profile of typical trench. (d) Full-width at half maximum (FWHM) measurements of trenches as a function of laser power and scan speed. (e) Trench depth as a function of laser power and scan speed

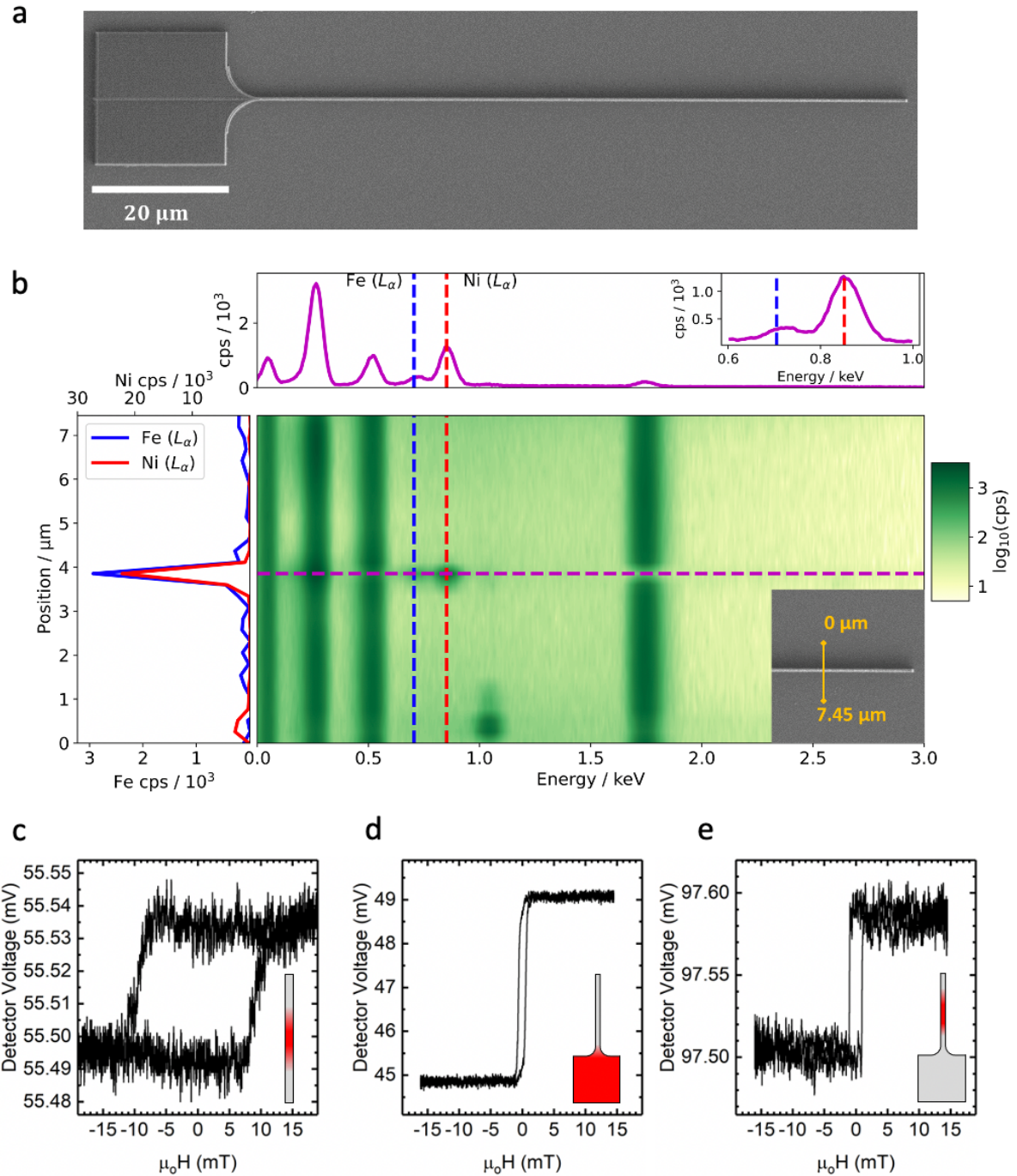

**Figure 3.** (a) SEM image of a 3D magnetic nanowire with a domain wall injection pad. (b) Spectrogram showing peaks for Ni, Fe and Na as a function of position. Inset: SEM showing section of nanowire from which spectrgram was obtained. (c) Magneto-optical Kerr effect (MOKE) loops obtained for single nanowire without domain wall injection pad. (d) MOKE loop obtained upon wire with injection pad, with laser focussed upon the pad. (e) MOKE loop obtained upon wire with injection pad, with laser focussed upon the wire.

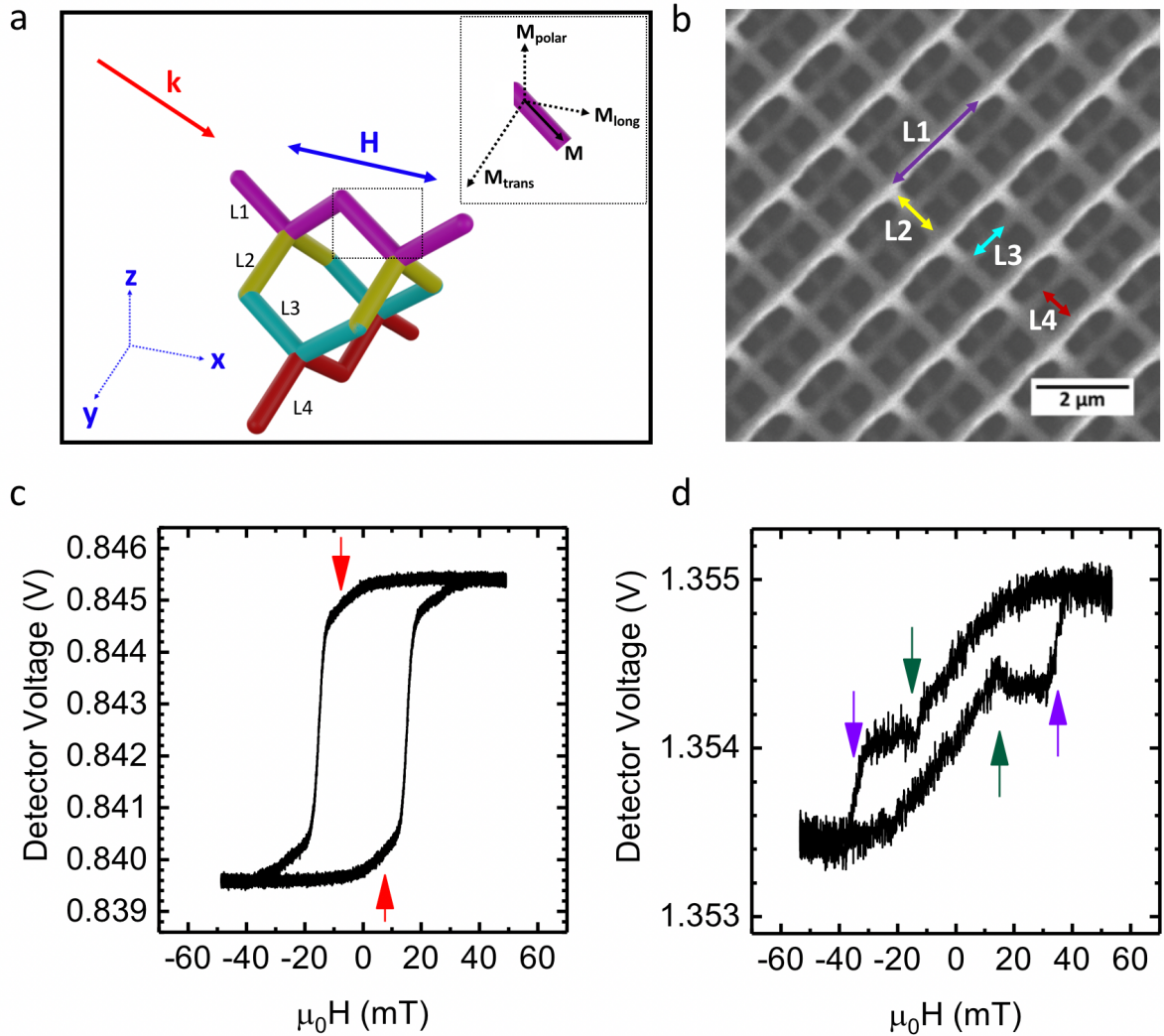

**Figure 4.** (a) Schematic of experimental magneto-optical Kerr effect (MOKE) geometry with 3D artficial spin-ice system. Unit cell of 3DASI is shown, with different sub-lattices depicted. The laser with wavevector k, makes an angle of 45 degrees with respect to the substrate and with projection along L1 as depicted. Inset: The magnetisation components that can be measured using the longituidnal, transverse and polar MOKE effects. (b) Scanning electron microscopy image of 3DASI structure taken at normal incidence. Different sub-lattices are labelled by colour. (c) MOKE loop taken using s-polarised light and with an analyser angle of 3 degrees. (d) MOKE loop taken using p-polarised light and with an analyser angle of 90 degrees.

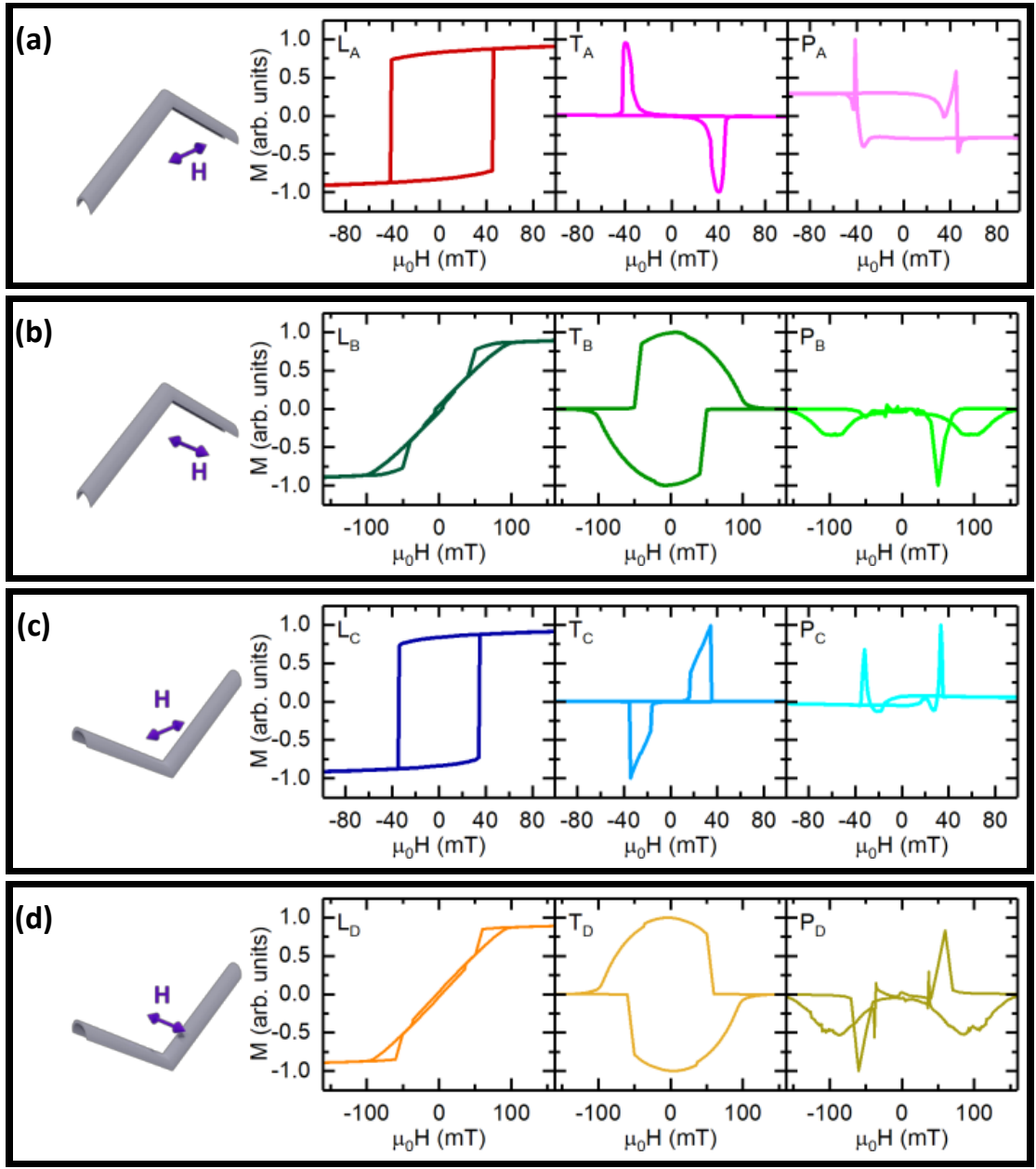

**Figure 5**. Micro-magnetic simulations obtained upon (a) upper bipod system, representative of 3DASI surface termination, with field along projection of long-axis, (b) upper bipod system, with field transverse to long-axis projection, (c) lower bipod system with field along projection of long-axis and (d) lower bipod system with field transverse to long-axis projection. For each simulation, the different components of magnetisation are plotted. $L_x$ represents component along field direction, $T_x$ represents component transverse to field direction and $P_x$ represents component perpendicular to the field direction.

**Table of contents figure**

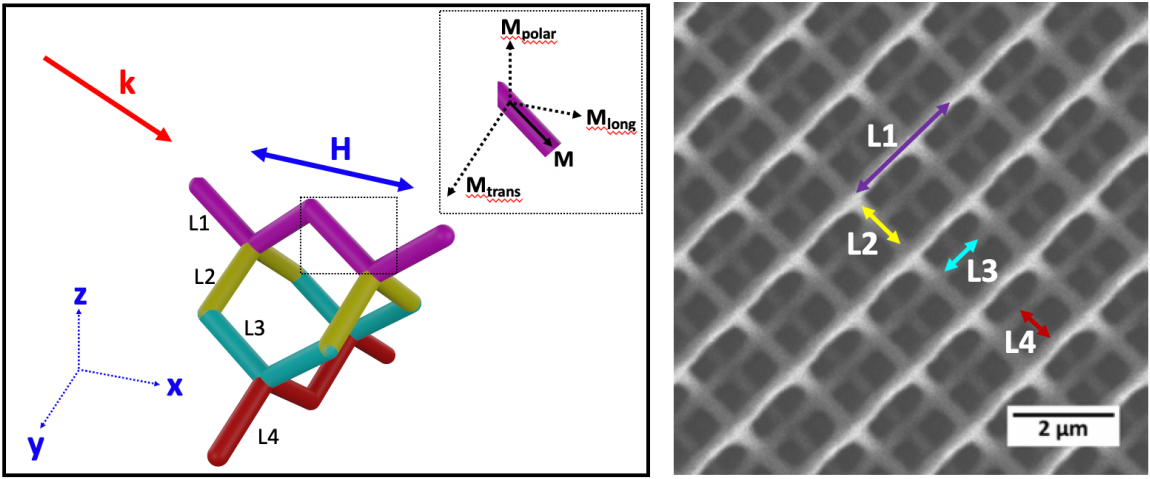